\documentclass{aa}
\usepackage{graphicx}
\usepackage{xcolor}
\usepackage{geometry}
\usepackage{xurl}
\usepackage[hidelinks]{hyperref}
\usepackage{natbib}
\usepackage{amsmath}
\usepackage{CJK}
\usepackage{appendix}
\usepackage{threeparttable}
\usepackage{amssymb}
\usepackage{lineno}
\usepackage{ulem}
\usepackage{caption}
\def\degree{${}^{\circ}$}

\linenumbers

\begin{document}

\title{A superflare of BP~Tau simultaneously caught by EP X-ray and TESS optical observations}

\author{Xingyu Zhou\inst{1,2}\fnmsep\thanks{Email: zhouxingyu@pric.org.cn} \and Mingjun Liu\inst{3} \and Gregory J. Herczeg\inst{4,5} \and P. Christian Schneider\inst{6,7} \and Fabio Favata\inst{8,9} \and Chenwei Yang\inst{1,2} \and Chichuan Jin\inst{3,10,11} \and Dongyue Li\inst{3} \and Xuan Mao\inst{3,10} \and Yi-Han Iris Yin\inst{12} \and Minghao Zhang\inst{3,10} \and Weimin Yuan\inst{3,10} \and Hongyan Zhou\inst{1,2,13}
        }

\institute{Polar Research Institute of China, 451 Jinqiao Road, Pudong, Shanghai 200136, People's Republic of China        
        \and Key Laboratory for Polar Science, MNR, Polar Research Institute of China, Shanghai, 200136, People's Republic of China
        \and National Astronomical Observatories, Chinese Academy of Sciences, 20A Datun Road, Beijing 100101, People's Republic of China
        \and Kavli Institute for Astronomy and Astrophysics, Peking University, Beijing 100871, People's Republic of China
        \and Department of Astronomy, Peking University, Beijing 100871, People's Republic of China
        \and Hamburger Sternwarte, Gojenbergsweg 112, D-21029, Hamburg, Germany
        \and Institut für Theoretische Physik und Astrophysik, Christian-Albrechts-Universität zu Kiel, Leibnizstrasse 15, 24118 Kiel, Germany
        \and INAF—Osservatorio Astronomico di Palermo, Piazza del Parlamento, 1, 90134 Palermo, Italy
        \and Department of Physics, Imperial College London, Exhibition Road, London SW7 2AZ, UK
        \and School of Astronomy and Space Science, University of Chinese Academy of Sciences, Chinese Academy of Sciences, Beijing 100049, People’s Republic of China
        \and Institute for Frontiers in Astronomy and Astrophysics, Beijing Normal University, Beijing 102206, People’s Republic of China
        \and Department of Physics, the University of Hong Kong, Pokfulam Road, Hong Kong, People’s Republic of China
        \and Key Laboratory for Research in Galaxies and Cosmology of Chinese Academy of Sciences, Department of Astronomy, University of Science and Technology of China, Hefei, People’s Republic of China}

\abstract{Multiwavelength observations of stellar flares trace the activity of different components of the stars' outer atmosphere, providing insight into their interactions. In the present paper, we report a superflare from BP~Tau, simultaneously observed with the Wide-field X-ray Telescope (WXT) on board the Einstein Probe (EP) satellite and TESS. While we attribute the X-ray flux increase to a magnetically powered flare, the optical light curve likely results from the superposition of the flare and an accretion burst. The X-ray flare has {a mean flux of $(1.5^{+0.3}_{-0.4})\times10^{-11}$ erg cm$^{-2}$ s$^{-1}$ in the WXT energy band (0.5-4.0 keV), with} e-folding times of $1.7\pm1.0$ ks and $14\pm5$ ks for the rise and decay phase, respectively. {The corresponding time-integrated} flare energy is $(1.0\pm 0.2)\times 10^{36}$ erg. The optical flare has an e-folding time of $0.33\pm0.04$ ks for the rise phase, but the data do not constrain the decay timescale. Assuming a decay phase equal to the rise phase, the resulting optical flare energy is $(2.8\pm0.4)\times10^{34}$ erg in the TESS band ($\sim6,000$-$\sim10,000$ \AA), corresponding to a bolometric energy of $(1.9\pm0.3)\times10^{35}$\,erg (assuming a blackbody at 11000\,K). The Follow-up X-ray Telescope (FXT) on EP triggered an observation $\sim1.5$ day after the flare, with a flux of $(4.6^{+0.2}_{-0.5})\times10^{-13}$ erg cm$^{-2}$ s$^{-1}$ (0.5-10.0 keV), indicating that BP~Tau had returned to quiescence. This work demonstrates the potential of jointly analyzing EP and TESS data for superflares. WXT is expected to detect $\sim800$ superflares per year, with FXT capable of slewing to the flaring star within $\sim3$--5 minutes. The large field of view of both missions offers us the opportunity to study multiwavelength variability during energetic flares.}

\keywords{stellar flares -- classical T-Tauri stars -- protoplanetary disks}

\maketitle
\nolinenumbers

\section{Introduction}
\label{sec:intro}

T-Tauri stars exhibit stronger stellar magnetic fields and more violent magnetic activities than main-sequence solar-type stars due to their larger convective zones and faster rotation. This results in X-ray luminosities several orders of magnitude higher than those of main-sequence stars \citep{Wright2011}. {X-ray flares are characterized by sharp brightening and an initially fast decay, followed by a much slower decay in light curves \citep{Favata2005, Stauffer2014}. These flares share similar heating and cooling processes with pre-main-sequence stars and solar-type stars \citep{Getman2021b}, while large flares are much more frequent in pre-main-sequence stars} \citep{Getman2021a}. 

{Classical T-Tauri stars, characterized by an accretion disk, have magnetic field lines truncating} the disk near the stellar surface. {Accreting material} flows from the inner disk to the star along magnetic field lines and hits the stellar surface at a near free-fall velocity (e.g., \citealt{Koenigl1991}, \citealt{Shu1994}; see also review by \citealt{Hartmann2016}). The impact generates X-ray emission mainly below 1 keV \citep{Lamzin1999}. The low energy component has been diagnosed via the plasma density derived from line ratios of the O VII and Ne IX triplets in high-resolution grating spectra (e.g., \citealt{Kastner2002}, \citealt{Argiroffi2007}, \citealt{Brickhouse2010}, \citealt{Gunther2013}; see also review by \citealt{Schneider2022}) and X-ray soft excess. In classical T-Tauri stars, the flux ratio between the O VII triplet and the O VIII {Lyman-alpha (Ly-$\alpha$)} lines is higher than in their diskless counterparts, weak-lined T-Tauri stars \citep[e.g.,][]{Gudel2007_soft_excess, Robrade2007, Telleschi2007_soft_excess, Brickhouse2010}. 

{The accretion process in classical T-Tauri stars is unstable, and light curves are considered as effective tracers of accretion variability. Early investigations, based on monitoring at cadences of days to years, attributed irregular photometric variations in classical T-Tauri stars to the accretion process \citep[e.g.,][]{Herbst1994, Lamm2004}. More recently, space telescopes including the Convection, Rotation, and planetary Transits \citep[CoRoT;][]{Baglin2009_CoRoT}, {Kepler} K2 \citep{Howell2014_K2}, and Transiting Exoplanet Survey Satellite (TESS; \citealp{Ricker2015_TESS}) have enabled high cadence photometric monitoring. Based on these data, bursts and some quasiperiodic signals in light curves of classical T-Tauri stars are attributed to accretion variability, with typical timescales ranging from hours to days (e.g., \citealp{Alencar2010, Cody2014, Cody2022, Stauffer2014}; see also review by \citealp{Fischer2023}). The accretion process has also been diagnosed via simultaneous multiwavelength monitoring and spectroscopic observations. Photometric fluxes correlate with accretion rates, as traced by either U-band flux excess over the photospheric level or the ratio of accretion flux to photospheric flux \citep[e.g.,][]{Robinson2022, Wendeborn2024b, Ji2026}. The correlation is not only common to classical T-Tauri stars in general but also present in our target, BP Tau \citep{Robinson2022, Wendeborn2024b}. These observations, including the light curve of BP Tau, are consistent with simulations of accretion disks where changes in the accretion rate result in variability in light curves \citep{Robinson2021, Robinson2022, Romanova2025}. More results specifically on BP Tau are introduced in the following context.}

\begin{figure*}[ht]
    \centering
    \includegraphics[width=0.8\hsize]{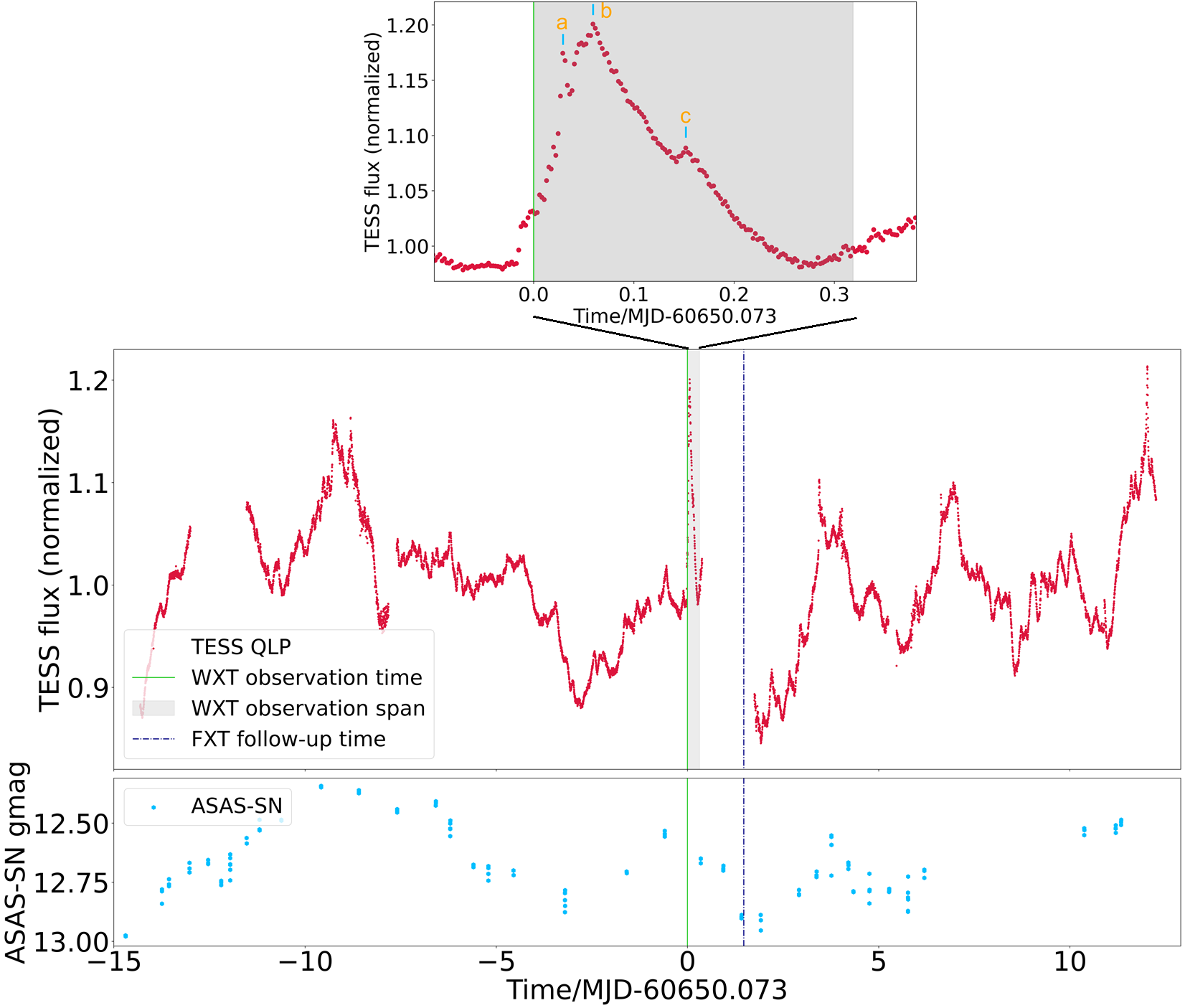}
    \caption{Bottom panels: TESS Sector 86 Quick-Look Pipeline light curve (top) and ASAS-SN {g}-band light curve (bottom) of BP~Tau. The vertical lines indicate the WXT observation time of the flare and the follow-up by FXT, with shadows covering the duration of the WXT observation. Top panel: TESS light curve zoom-in on the flare. The short blue line segments indicate the peaks of the TESS light curve, {labeled as ''a,'' ''b,'' and ''c'' in chronological order.}}
    \label{fig:lc_ASASSN_TESS_EP}
\end{figure*}

In classical T-Tauri stars, flaring magnetic loops can reach the accretion disk, and strong flares may perturb the disk and trigger accretion bursts, as investigated and suggested by both observations and simulations. An early study by \citet{Favata2005} identified intense flares in classical T-Tauri stars with loop heights far exceeding stellar radii, potentially reaching the inner edge of the accretion disk. A follow-up study by \citet{Aarnio2010} estimated the disk inner radii of the sample in \citet{Favata2005} and found that up to $\sim40\%$ of the sample have flare loop heights consistent with those of the inner edge of their disks. This pointed to potential connections between flares and disks. \citet{Reale2018} present a different method of estimating the loop heights of flares, showing that the loop heights may reach the inner disk and perturb accretion. Indirect observational evidence of a flare-triggered accretion burst comes from a simultaneous {HST}-{Swift}-{Chandra} observation of several classical T-Tauri stars, where a correlation between the accretion column energy flux and X-ray luminosity is found \citep{Espaillat2019}. Recent large-sample studies of well-defined {superflares (total energy $>10^{34}$ erg)} show that {superflares exhibit similar properties, such as energy distributions and plasma temperatures in both classical T-Tauri stars and weak-lined T-Tauri stars.} Statistically, this suggests no evidence for a disk-related flare mechanism \citep{Getman2021a, Getman2021b, Getman2024}. The inverse process, in which flares trigger accretion bursts, remains a possibility \citep{Getman2021a}. Some simulations show that energetic flares can create large loops connecting the star to the accretion disk, perturbing its stability. This may trigger an enhanced accretion rate and predict a timescale of several hours to one day between the flare and the resulting accretion enhancement \citep[e.g.,][]{Orlando2011, Colombo2019}. The simulation by \citet{Waterfall2019} assumes that flares are associated with accretion disks. It reproduced the observed X-ray and radio luminosities of flares.

While stellar coronal activity has in the past been investigated through high-energy emission (X-rays and UV data), simultaneous multiwavelength observations make it possible to observe the interaction between different components of a star's outer atmosphere -- jointly, in the case of young accreting stars, with accretion processes (e.g., as traced by H$\alpha$). Such observations have enabled the study of accretion-related X-ray luminosity variations \citep[e.g.,][]{Dupree2012, Bustamante2016, Guarcello2017, Schneider2018, Espaillat2019}, stellar surface magnetic fields following large X-ray flares \citep{Getman2024, Getman2025}, as well as thermal and nonthermal emission in flares (e.g., the Neupert effect, \citealt{Neupert1968}; see observations in \citealt{Gudel2004} and \citealt{Stelzer2022b}).

The aforementioned multiwavelength observations were carried out by aligning the observation schedules of different telescopes. Another approach is to search for overlapping periods when two sky-survey telescopes coincidentally observe the same target. This was achieved, for example, using the extended Roentgen Survey with an Imaging Telescope Array (eROSITA; \citealp{Predehl2021_eROSITA}) and its wide field of view (1.03\degree diameter per module, with seven modules in total), along with TESS (24\degree$\times$96\degree field of view), as demonstrated by the simultaneous flare observation of M dwarfs in \citet{Joseph2024}. Given the stochastic nature of flares -- in particular, rare super- or mega-flares \citep[e.g.,][]{Getman2021a} -- such a method benefits from telescopes with larger fields of view.

Launched in 2024, the Einstein Probe Mission \citep[EP;][]{Yuan2022_EP} carries a {Wide-field X-ray Telescope}  (WXT) with a field of view of $\sim3600$ square degrees and a detection limit of $\sim2.6\times10^{-11}$ erg s$^{-1}$ cm$^{-2}$ at 1 ks\footnote{0.5-4.0 keV, assuming a power-law spectrum with a photon index of 2 and a Galactic absorption of $3\times10^{20}$ cm$^{-2}$} \citep{Yuan2025_EP}, providing great potential for the study of energetic flares \citep[e.g.,][]{Gunther2024, Mao2025}\footnote{The flare in these two studies was detected by the Lobster Eye Imager for Astronomy \citep{Zhang2022_LEIA, Ling2023_LEIA}, the pathfinder of EP.}. EP also carries a {Follow-up X-ray Telescope} (FXT), which achieves deeper sensitivity ($\sim1\times10^{-14}$ erg s$^{-1}$ cm$^{-2}$ at 10 ks; see \citealp{Zhang2022_FXT} for a detailed estimate) and is designed for the quick follow-up (as fast as 3-5 minutes) of X-ray events detected by WXT, as well as target-of-opportunity (ToO) observations \citep{Yuan2025_EP}.

Here, we report a superflare from BP~Tau simultaneously detected by WXT and TESS, with a follow-up by FXT $\sim$\,1.5\,days after the flare. BP~Tau is an M0.5 classical T-Tauri star \citep{Herczeg2014} located at 128\,pc (based on the {Gaia} {Data Release 3 (DR3)} parallax, \citealp{Gaia_Mission, GaiaDR3}). The X-ray emission of BP~Tau can be described by a thermal spectrum with three characteristic temperatures. The hotter two ($\sim$0.5 and $\sim$2.1\,keV) dominate the X-ray emission and are likely of coronal origin, while the coolest one ($\sim0.2$ keV) is partially produced via the accretion process \citep{Schmitt2005, Robrade2006}. {X-ray flares of BP~Tau were observed by {Chandra} (ObsIDs 16205 and 16558) and {XMM-Newton} (ObsIDs 0200370101 and 0882060801), with peak count rates increasing by a factor of $\sim8$ (two {Chandra} flares) and $\sim4$ (two {XMM-Newton} flares).} Several optical flares of BP~Tau have been detected, with energies ranging from $10^{34}$ to $10^{35}$ erg in the TESS band \citep{Lin2023}.

{BP~Tau is an extensively studied target for investigating the accretion process in classical T-Tauri stars and its resulting photometric variability. Simultaneous multiband observations detected brightening events with nearly symmetric light curve profiles and temperatures of 7000-8000 K at peak fluxes attributed to inhomogeneous accretion rather than flares \citep{Gullbring1996a}. This was further supported by simultaneous optical and X-ray observations of BP Tau, in which no correlation between optical and X-ray fluxes was found \citep{Gullbring1997}. Multiple spectropolarimetry observations of BP Tau by \citet{Donati2008} revealed broad emission lines associated with accretion (H$\alpha$, H$\beta$, the broad component of the Ca II IR triplet) that varied beyond rotational modulation. As an explanation, \citet{Costigan2014} suggested that the H$\alpha$ emission arises from the bulk of accretion streams and is therefore more sensitive to accretion instability. \citet{Donati2008} also detected optical veiling, which weakens photospheric absorption lines. They attributed veiling variations to unstable accretion rates.} {High cadence} optical light curves of BP~Tau exhibit strong variability with bursts and dips \citep[e.g.,][]{Lin2023, Burlak2025}, {which is attributed to its variable accretion rate \citep{Robinson2022, Wendeborn2024b}.} {Spectroscopic monitoring of BP~Tau shows that its accretion rate varies peak to peak by a factor of 3.6 across 21 observations \citep{Wendeborn2024a}.} 

The present paper is organized as follows. We describe data reduction process of both EP and TESS in Sect. \ref{sec:data}. We present an analysis of the data and results in Sect. \ref{sec:flare_analysis} and discuss them in Sect. \ref{sec:discussion}. We summarize our work in Sect. \ref{sec:conclusion}.

\begin{figure}[t]    
    \centering
    \includegraphics[width=\hsize]{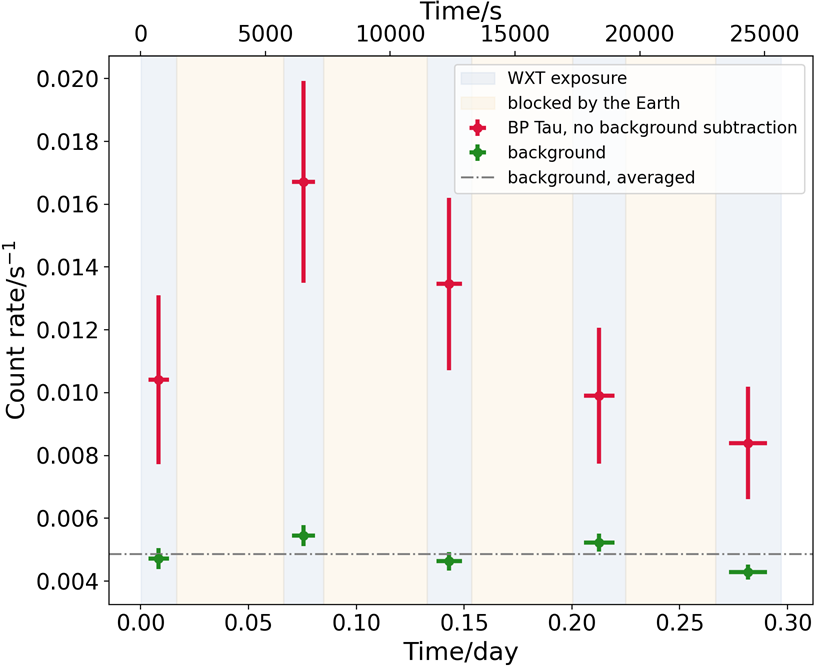}
    \caption{WXT light curve of BP~Tau (top) and the corresponding background (bottom), binned in each uninterrupted observation. {The five segments have exposure times of 1441 s, 1616 s, 1783 s, 2121 s, and 2621 s, respectively, and are indicated by the light blue areas}.}
    \label{fig:lc_WXT_check_bkg}
\end{figure}

\section{X-ray and optical data}
\label{sec:data}

The X-ray flare of BP~Tau was detected by WXT on UTC 2024-12-06 01:45:01 (MJD=60650.073). The observation spans 25.65 ks, with a total exposure time of 9.66 ks and the remainder blocked by the Earth {or influenced by the South Atlantic Anomaly region}. {The good time intervals of this observation were split into six segments with exposure times of 1441 s, 79 s, 1616 s, 1783 s, 2121 s, and 2621 s, respectively. We do not discuss the segment with 79 s exposure time further and focus our analysis solely on the five long segments.} Later, FXT performed a follow-up observation on UTC 2024-12-07 13:04:44 (MJD=60651.545), a ToO observation triggered by on-ground data processing. {An observation table of WXT and FXT is presented in Table \ref{table:EP_obs_table}. The energy ranges of WXT and FXT are 0.5-4.0 keV and 0.3-10.0 keV, respectively \citep{Yuan2025_EP} } 

Meanwhile, BP~Tau was covered by TESS Sector 86 from 2024-11-21 to 2024-12-18. The TESS light curve {generated by the Quick-Look Pipeline \citep[QLP;][]{Huang2020_QLP, Huang2020b_QLP, Kunimoto2021_QLP, Kunimoto2022_QLP}}, together with the WXT and FXT observations, is shown in Fig. \ref{fig:lc_ASASSN_TESS_EP}. {Data points affected by the gap between TESS detectors were visually identified and removed.} We also include the {g}-band ($\sim4,000$-5,000\AA, i.e., bluer than the TESS band) light curve from the ASAS-SN \citep{Shappee2014_ASASSN, Kochanek2017_ASASSN}. This enabled us to assess the flux variation when BP~Tau lay in the gap between TESS detectors. The wavelength range of the TESS band is $\sim6,000$-$10,000$\AA\footnote{See \url{https://heasarc.gsfc.nasa.gov/docs/tess/telescope\_information.html} for the TESS response function}. The data processing of EP and TESS is described below.

\begin{table}[t]
    \caption{WXT and FXT observation table.}
    \label{table:EP_obs_table}
    \centering    
    \begin{threeparttable}
    \begin{tabular}{lllll}
    \hline
    \hline
           & WXT   & FXT   \\ \hline
    Obs ID &  11904195783  &  06800000280   \\
    Start (UTC)   &  2024-12-06 01:45:01   &  2024-12-07 13:04:44   \\
    On-source &  25.65 ks  &  3.1 ks  \\
    Exposure &  9.66 ks  &  3.1 ks  \\ \hline
    \end{tabular}
    \end{threeparttable}
\end{table}

\subsection{EP}
\label{subsec:data_EP}
The position of the WXT source is {(64.838\degree, 29.123\degree) in Right Ascension and Declination} ($2.5$\arcmin error circle at 90\% confidence level), $1.5$\arcmin\ away from BP~Tau. This discrepancy is consistent with the $\leq5$\arcmin\ typical angular resolution of WXT. To rule out contaminating sources, we used {Gaia} DR3 and XEST \citep{Gudel2007_XEST} data, searched for sources in the two catalogs within the $2.5$\arcmin\ error circle of the WXT position, and cross-matched the sources in a radius of $3$\arcsec. While BP~Tau was the only remaining source, we also considered the possibility of stars without previous X-ray detection that might have been flaring in this observation. {Gaia} DR3 shows that within the error circle, BP~Tau is the only star within 250 pc. Given the observed flux in the WXT, a flared star would need to be sufficiently close to ensure that the flare luminosity remains within a reasonable range (e.g., several times $10^{32}$ erg s$^{-1}$). Considering the X-ray activity and distance of BP~Tau, a flare in BP~Tau is the most probable source of the X-ray detection. 

WXT data were reduced using the WXT Data Analysis Software developed for the EP mission (Liu et al., in prep.) and the first version of the calibration database (CALDB) generated for WXT \citep{Cheng2025_EP}. During the $\sim7$\,h observation of BP~Tau, WXT was blocked by the Earth four times. In each uninterrupted observation segment, the count rate does not exhibit measurable variations. We binned the photons in each segment into a single data point to increase the signal-to-noise ratio, resulting in a light curve with five data points, as shown in Fig. \ref{fig:lc_WXT_check_bkg}. The light curve of BP~Tau exhibits count rate variability and is different from the light curve of the background, ruling out background fluctuations as the cause of variability. The background-subtracted light curve in the following analysis (Fig. \ref{fig:alignment_WXT_TESS}) incorporates the averaged background of the 5 epochs.

The follow-up of BP~Tau by FXT is continuous in its $\sim3000$\,s observation and shows no significant variability. FXT data were reduced using the data reduction software developed and provided by the EP Science Center with the latest FXT calibration database (CALDB v1.10).

\subsection{TESS}
\label{subsec:data_TESS}

\begin{figure}[t]
    \centering
    \includegraphics[width=\hsize]{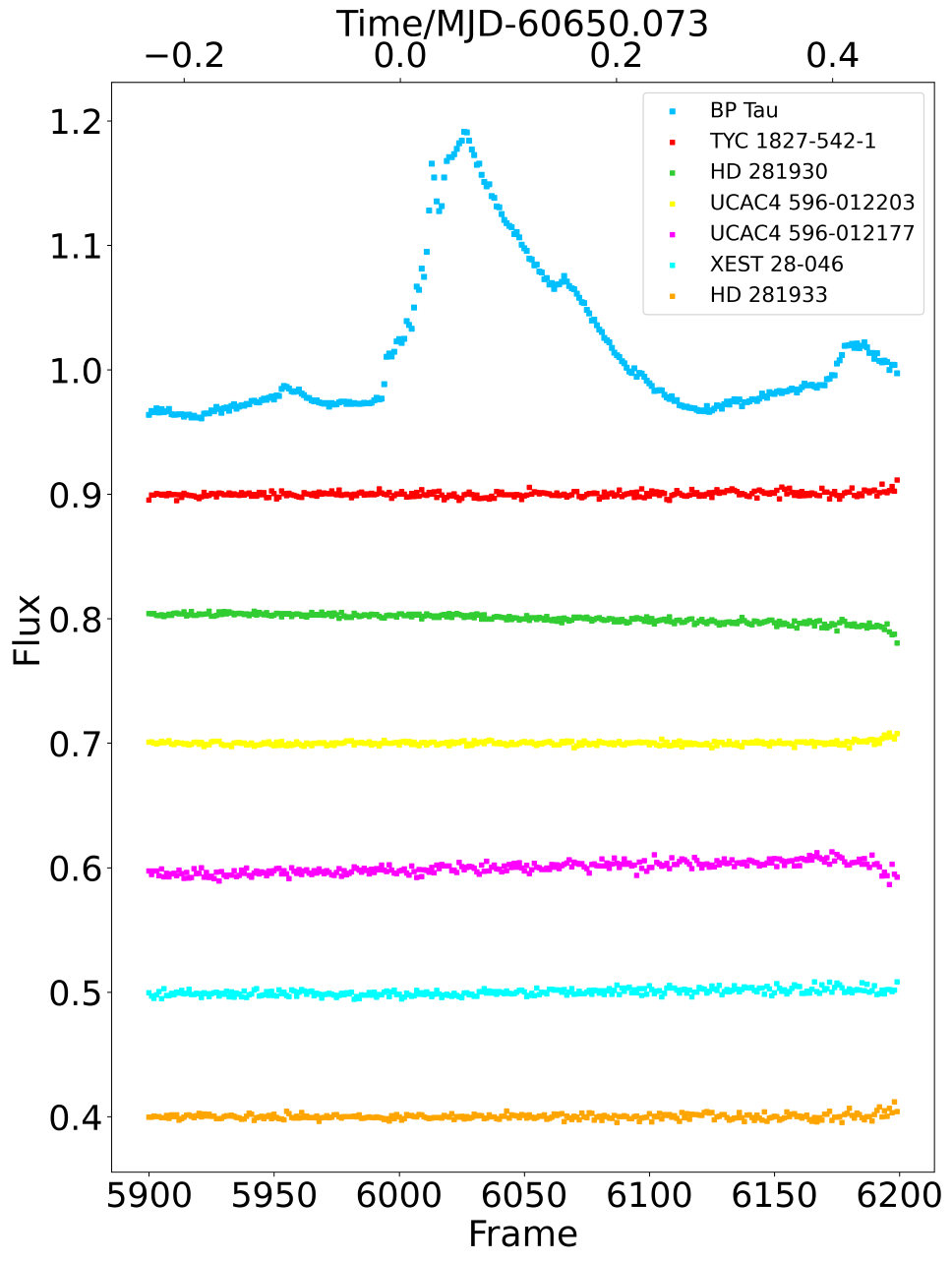}
    \caption{{Differential photometry light curves of BP~Tau and six comparison stars. The WXT observation toward BP~Tau covers 0.00-0.32 in the ''Time'' axis. All the light curves are normalized, and the light curves of the comparison stars are vertically shifted.} One frame on the x-axis is 200\,s.}
    \label{fig:TESS_BPTau_nearby_star}
\end{figure}

{Our differential photometry result is used in the following analysis of the flux bump covering the WXT observation ($\sim-0.1<\mathrm{MJD}-60650.073<\sim0.3$). We performed differential photometry via AstroImageJ \citep{Collins2017_AstroImageJ} on the TESS full-frame image cutouts generated by TESSCut \citep{Brasseur2019_TESSCut}. Comparison stars in differential photometry were selected from stars within 10' of BP~Tau and with TESS Tmag within 1 mag of BP~Tau (according to the TESS Input Catalog version 8.2, \citealp{Paegert2021_TIC}), yielding a total of six stars: TYC 1827-542-1, UCAC4 596-012177, XEST 28-046, HD 281930, UCAC4 596-012203, and HD 281933.} Apertures and backgrounds were visually selected pixel by pixel to avoid contamination from other stars. Sources within $2$\arcmin\ of BP~Tau from the {Gaia} DR3 catalog are all at least 3 mag fainter in the G band, providing no contamination to the light curve of BP~Tau at the $21$\arcsec\ pixel scale of TESS. The resulting light curve is presented in Fig. \ref{fig:TESS_BPTau_nearby_star}.

{Since BP~Tau moved into the gap between TESS detectors $\sim0.1$ day after its flux bump under analysis, as shown in Fig. \ref{fig:lc_ASASSN_TESS_EP}, we assessed potential influence of the gap on the light curve of the flux bump. Background fluxes of BP~Tau and its comparison stars in differential photometry increase smoothly during the flux bump with nearly identical increase rates, indicating no significant jumps in the background flux and no obvious background gradients within each frame. This is consistent with our visual inspection of images frame-by-frame. As the background brightens, the signal-to-noise ratio of BP~Tau decreases from 65 to 57, which nevertheless provides reliable photometric results. Hence, gap influence is not considered in the following analysis.}

\section{Analysis of X-ray and optical flare}
\label{sec:flare_analysis}

\begin{figure}[t]
    \centering
    \includegraphics[width=\hsize]{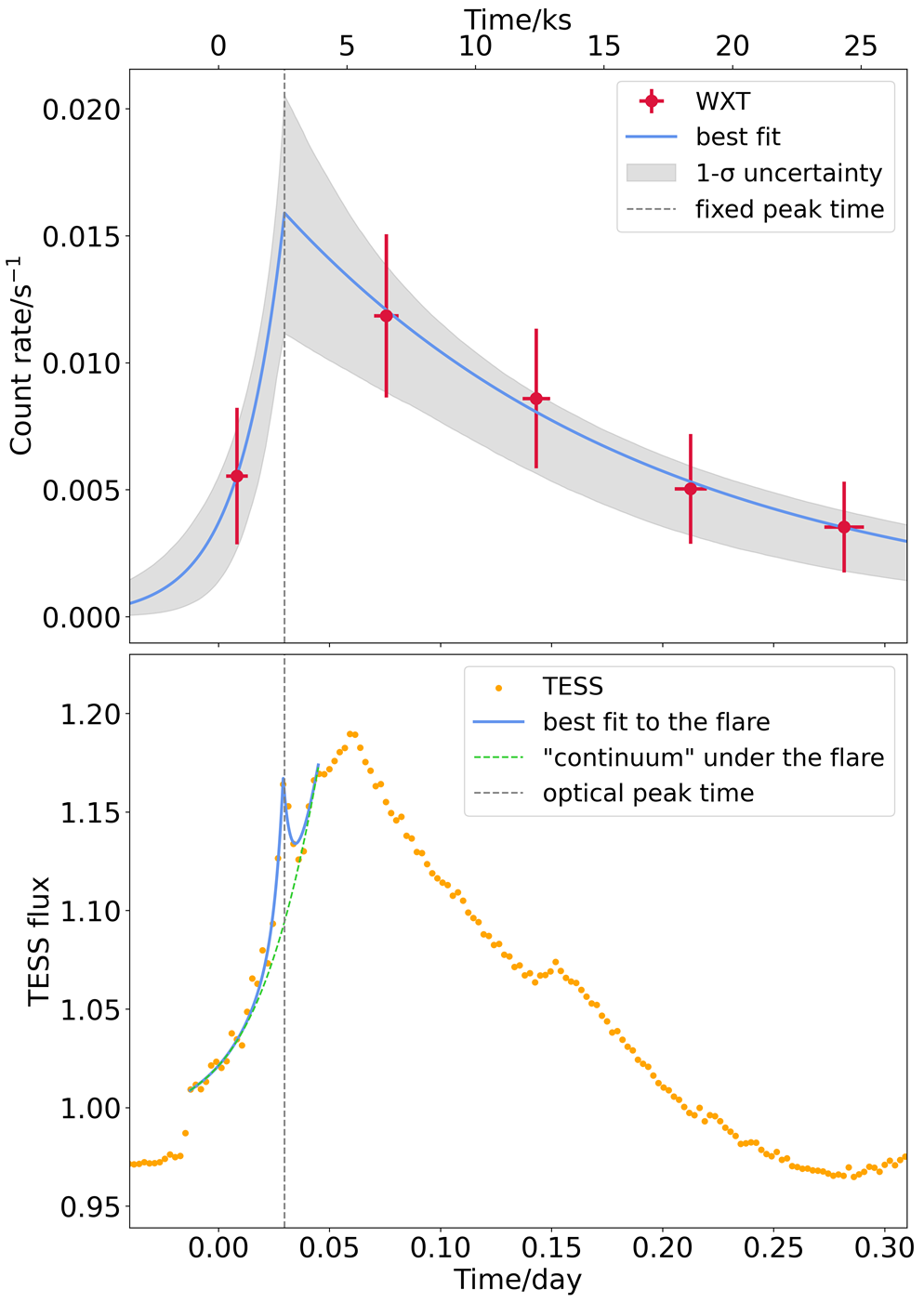}
    \caption{Alignment of the WXT (top) and TESS (bottom) light curves of BP~Tau, together with the best-fit results for the flare in blue. In the TESS light curve, the dashed green curve shows the {slow modulation baseline}, and the blue curve represents the {flare above the baseline}. The zero point in the x-axis corresponds to MJD=60650.073, which is the time at the WXT observation began. The vertical dashed line represents the optical flare peak time -- assumed to be the same as the X-ray flare peak time and used for fitting the X-ray light curve. Our fitting procedure is described in detail in Sect. \ref{subsec:flare_energy}.}
    \label{fig:alignment_WXT_TESS}
\end{figure}

\subsection{Interpretation of the light curves}
\label{subsec:lc_interpret}

Light curves from WXT and TESS are presented in alignment in Fig. \ref{fig:alignment_WXT_TESS}. {Despite its considerable uncertainties, the WXT light curve still exhibits flare-like variability, which is analyzed statistically in Sect. \ref{subsubsec:WXT_variability}}. The TESS light curve shows three peaks at $\sim0.03$, $\sim0.06$, and $\sim0.15$\,day, as shown in Fig. \ref{fig:lc_ASASSN_TESS_EP}, with the first one sharper than the rest. While the {flux bump between $\sim-0.1<\mathrm{MJD}-60650.073<\sim0.3$ in the TESS light curve} appears similar to the X-ray light curve, we suggest that {peak ''a'' in Fig. \ref{fig:lc_ASASSN_TESS_EP}} corresponds to an optical flare associated with the X-ray flare and that the longer bump between $\sim-0.1<\mathrm{MJD}-60650.073<\sim0.3$ is more likely associated with an accretion burst. This interpretation is supported by the following {analysis}.

\subsubsection{{Variability in the WXT light curve}}
\label{subsubsec:WXT_variability}

{Assuming that the peak at $\sim0.03$ day in the TESS light curve is the optical counterpart of the X-ray flare, we expect the WXT light curve to show a typical shape of an X-ray flare, with a peak near its optical counterpart and a much slower decay phase. Such expectation is indicated by flare models, where the optical emission is a result of plasma heating in the photosphere or chromosphere, and the soft X-ray emission originates from evaporated plasma lifted into the magnetic loops \citep{Kowalski2024}. }

Considering the uncertainties and limited sampling of the X-ray light curve, we employed simulations to quantitatively assess the presence of variability. {We fitted the observed light curve using both a constant model and the flare model in \citet{Getman2021b} and calculated the difference in their C-statistics ($\Delta C=C_{flare}-C_{const}$). During the flare model fitting, the ratio of the rise to decay e-folding timescales was fixed and determined from the local regression fit to the data in \citet{Getman2021b}. The light curve without background subtraction was utilized because its photon counts follow the Poisson distribution and a constant background count rate was added to the flare model. The resulting $\Delta C$ was $-6.9$. To assess its significance, we simulated 10,000 light curves based on a constant count rate (total counts in the 9.66 ks exposure divided by the exposure time) and Poisson errors. Each simulated light curve was then fitted with both models to obtain the distribution of $\Delta C$. Among these simulations, 2.4\% of the simulated $\Delta C$ values was less than the difference derived from the observation. Hence, there is a 2.4\% probability that the observed variation is a result of Poisson noise, providing 97.6\% confidence of rejecting the constant model.}

{Under the flare model, the second segment of the light curve is near the peak of the flare and is followed by a declining trend. Given this scenario, we compared the hardness ratio of the second segment with the overall hardness ratio of the complete WXT observation. The hardness ratio, defined as the ratio between the count rates in 1.25-4.0 and 0.5-1.25 keV, was calculated via the program Bayesian Estimation of Hardness Ratios \citep{Park2006_BEHR}, in which the counts from both the source and background are considered simultaneously. The second segment of the WXT light curve has an 81\% probability of being harder than the entire WXT observation. {This is} consistent with the scenario in which the increase in the count rate is accompanied by an increase in the hardness ratio, further supported by the presence of a harder component in the WXT energy spectrum (see Sect. \ref{subsec:flare_energy}).}

\subsubsection{{Decomposing the TESS light curve}}
\label{subsubsec:decompose_TESS}

{We first considered whether the light curve is a superposition of flares \citep[e.g., Fig. 6 of][]{Davenport2014}. Following the method in \citet{Davenport2014}, we iteratively fitted superpositions of multiple flare templates and compared the Bayesian information criterion (BIC) of each fit. We used the flare model from \citet{Mendoza2022}, which is an updated model by \citet{Davenport2014}. The uncertainty of TESS photometry was set to its typical photometric uncertainty, 60 ppm\footnote{See \url{https://heasarc.gsfc.nasa.gov/docs/tess/telescope\_information.html} for the uncertainty.}, when fitting and calculating the BIC.}

\begin{figure}[t]
    \centering
    \includegraphics[width=\hsize]{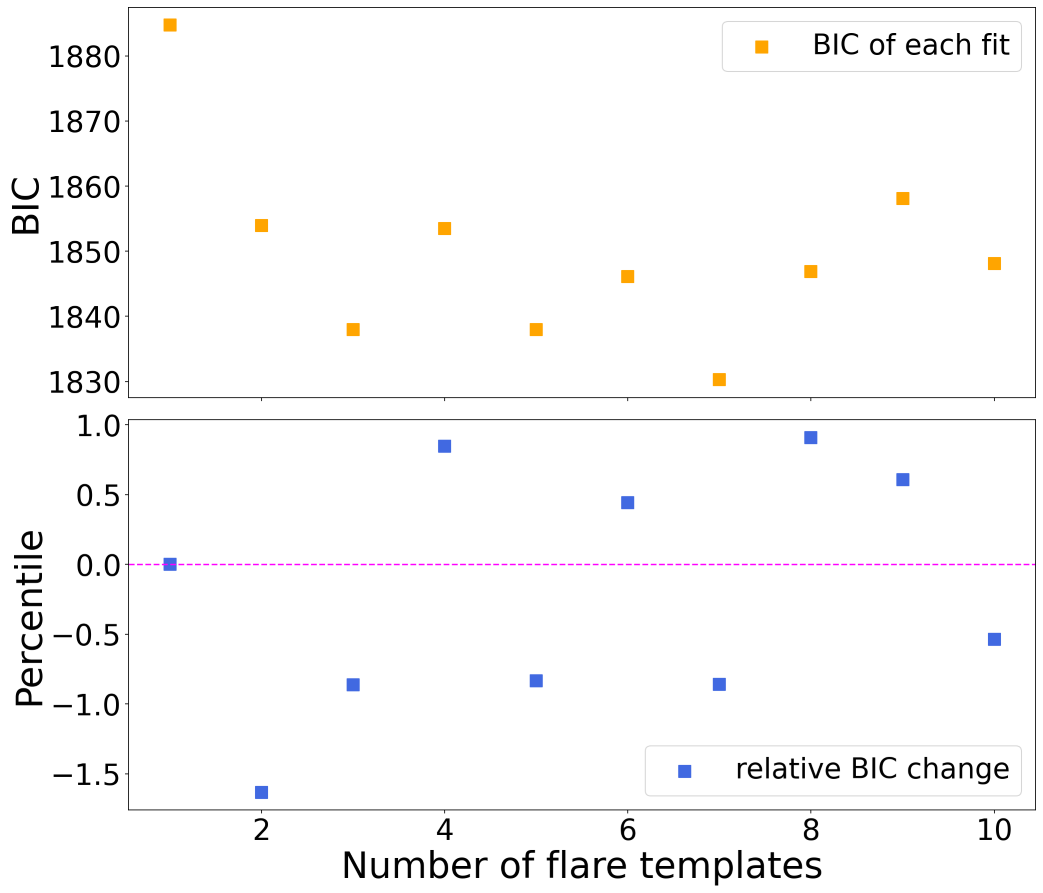}
    \caption{{Top: Relation between the number of stacked flare templates in the fitting and the corresponding BIC of the best-fit results. Bottom: Change in BIC (in percentiles) relative to the previous fit, $(\mathrm{BIC}_n-\mathrm{BIC}_{n-1})/\mathrm{BIC}_{n-1}$, where $n$ is the number of flare templates.}}
    \label{fig:check_BIC}
\end{figure}

\begin{figure}[t]
    \centering
    \includegraphics[width=\hsize]{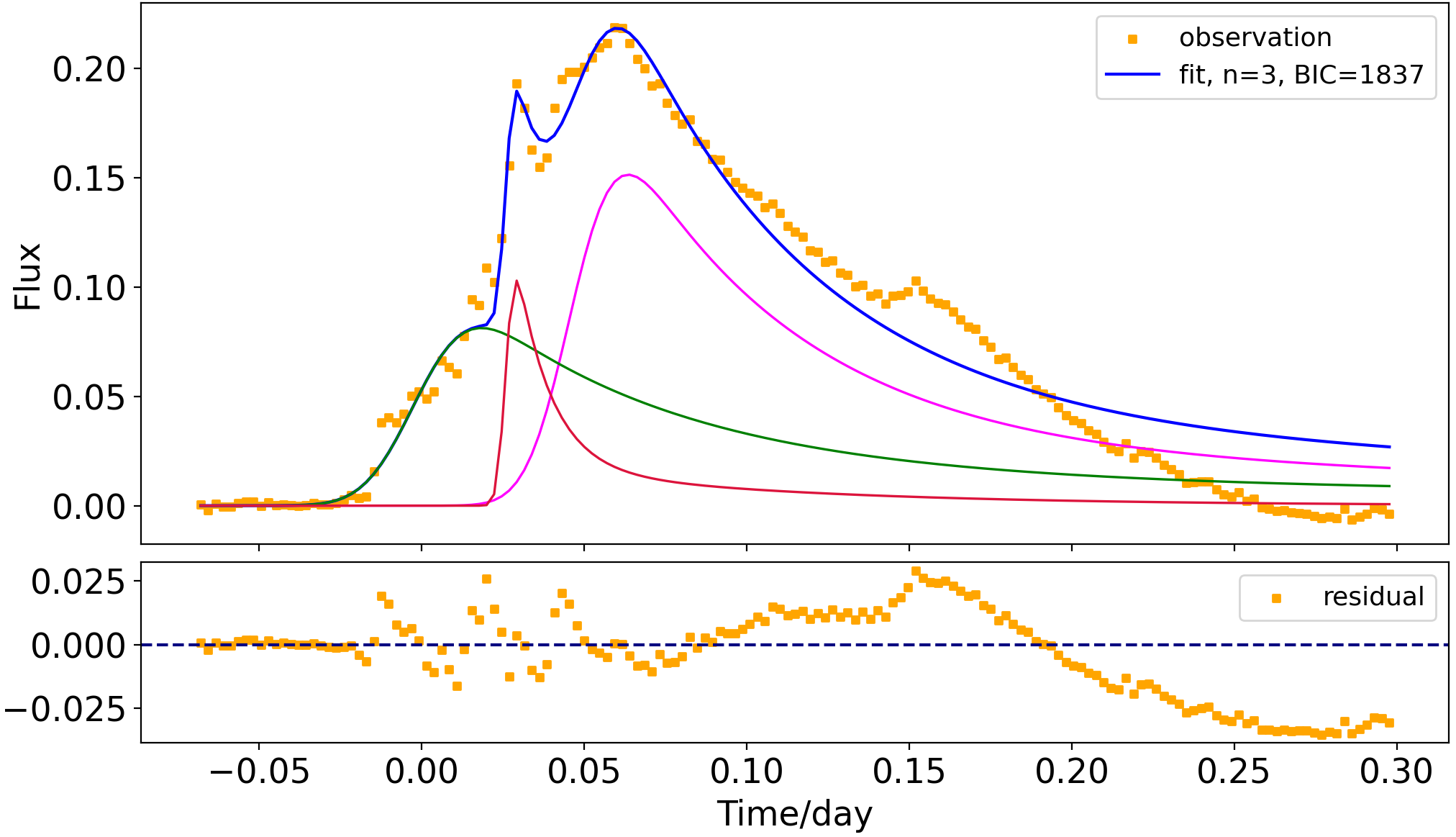}
    \caption{{Decomposing the TESS light curve with {three} flare templates.}}
    \label{fig:decompose_TESS_lc}
\end{figure}

{By increasing the number of flare templates from $n=1$ to $n=10$, we present the relation between $n$ and the corresponding BIC in Fig. \ref{fig:check_BIC}, together with the relative BIC change $(\mathrm{BIC}_n-\mathrm{BIC}_{n-1})/\mathrm{BIC}_{n-1}$. The most significant change in BIC occurs from n=1 to n=2, with a decrease in 1.6\%. Further increasing the number of templates does not significantly affect the BIC. {The best-fit results using three templates} is presented in Fig. \ref{fig:decompose_TESS_lc} {as an example}, which {does not provide} a satisfactory fit to the observation. The decay phase from the models is more gradual than that of the observation in the later segment ($>\sim0.18$ day) and steeper in the middle segment ($\sim0.10$-0.18 day), and does not reproduce peak ''c'' in Fig. \ref{fig:lc_ASASSN_TESS_EP}.}

{The decay phase of the TESS light curve was also fitted with four flare models from \citet{Yudovich2025}: the ''Duo-Classical'' model and the ''peak-bump'' models with chi-square, lognormal, and Gaussian bumps, respectively. Similar to the results presented above, the new fitting results also decay more gradually in the later segment and more steeply in the middle segment, indicating that the TESS light curve, particularly the decay phase, is not likely a superposition of flares.}

\subsubsection{{Similarity to accretion-driven variability}}
\label{subsubsec:compare_accretion_lc}

{Classical T-Tauri stars exhibit accretion variability, commonly traced by their irregular light curves \citep[e.g.,][]{Alencar2010, Cody2014, Wendeborn2024b, Ji2026}. Accretion bursts frequently occur in classical T-Tauri stars \citep[e.g., 0.2 bursts per day,][]{Stauffer2014} and have a timescale as short as several hours (see Fig.17 in \citealp{Stauffer2014}). We assessed whether the flux bump between $\sim-0.1<\mathrm{MJD}-60650.073<\sim0.3$ in the TESS light curve of BP~Tau results from accretion-burst-driven variability by comparing the light curve with accretion-burst-dominated light curves identified by \citet{Stauffer2014}.}

{The similarity was evaluated by the mean-square error and Pearson correlation coefficient between the two light curves. We first resampled the TESS light curve by linearly interpolating over data points to match the sampling cadence of CoRoT light curves (512 s). For each CoRoT light curve, segments were extracted by sliding a time window equal in length to the TESS light curve from the first point with a step size of one. The mean-square error and correlation coefficient were computed between each CoRoT segment and the TESS light curve. Both light curves were normalized by dividing by their respective mean fluxes prior to the calculation. }

\begin{figure}[t]
    \centering
    \includegraphics[width=\hsize]{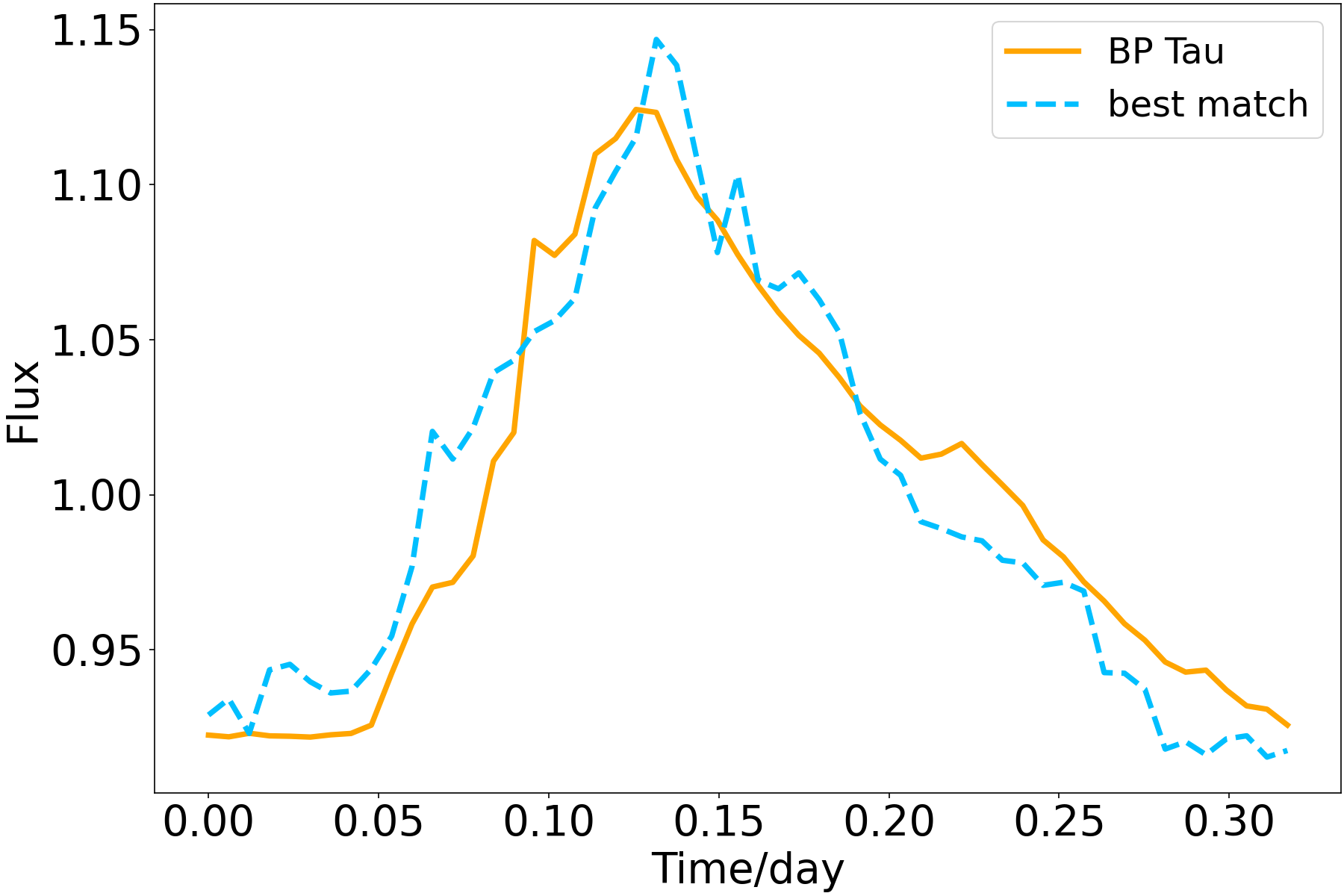}
    \caption{{{Segment} in Mon 469's CoRoT light curve most similar to the TESS light curve of BP~Tau.}}
    \label{fig:compare_BPTau_Mon469}
\end{figure}

{The {best match} between the CoRoT light curve of Mon 469 and the TESS flux bump according to the mean-square error (0.000429) is shown in Fig. \ref{fig:compare_BPTau_Mon469}. The corresponding Pearson correlation coefficient is 0.9503. Both similarity criteria indicate that the segment closely resembles the TESS light curve, suggesting that the variability of BP~Tau is caused by an accretion burst. Meanwhile, this similarity comparison is insensitive to local small structures; hence, the comparison result does not contradict our initial interpretation that the TESS light curve is composed of both an accretion burst and a flare.}

\subsubsection{{A flare in the TESS light curve}}
\label{subsubsec:TESS_flare}

{The aforementioned analysis indicates that the TESS light curve covering the WXT observation ($\sim-0.1<\mathrm{MJD}-60650.073<\sim0.3$) is more consistent with an accretion origin than with stellar flares. Meanwhile, } the X-ray emission from accretion bursts exhibits a low temperature (less than $\sim1$ keV; \citealt[e.g.,][]{Lamzin1999, Robrade2006}), at the lower bound of the 0.5-4.0 keV energy range of WXT and has been observed to be $\sim2-3$ times brighter than the quiescent flux (e.g., see time variability of X-ray properties in \citealp{Guarcello2017}). This is much fainter than the {WXT flux} ($\sim25$ times the quiescent flux of BP~Tau; see Sect. \ref{subsec:flare_energy} for the estimate). Hence, the X-ray flux seen in the WXT observation likely results from magnetic activity, and the (highly) elevated flux level shows that it must be from a flare.

{Considering that stellar flares can occasionally be detected simultaneously in both X-ray and optical observations, we suggest that {peak ''a''}  in the TESS light curve most likely represents the optical counterpart to the X-ray flare based on the variability timescale.} Timescales of optical flares are typically a few to tens of minutes and unlikely to exceed $\sim0.2$ day \citep[e.g.,][]{Hawley2014, Stauffer2014, Yang2023_flare}. {Peak ''a''} has a timescale consistent with an optical flare (see the {two independent fits} to the peak in Sect. \ref{subsec:flare_energy} {and in Fig. \ref{fig:decompose_TESS_lc}}). The timescale of X-ray superflares is typically hundreds of minutes long \citep{Getman2021a}, and simultaneous observations in X-ray and optical show that X-ray flares are significantly longer {than their optical counterparts} \citep[e.g.,][]{Stelzer2022b}.

{Therefore, as a summary of the previous analysis,} we adopted the interpretation of the {TESS} light curve as the superposition of a short flare and a longer accretion event and acknowledge the possibility of misinterpretation considering the small number of data points in the optical flare, the mixture of various peaks in the {TESS light curve}, and the discontinuity in WXT data.

\subsection{{Observations after the flare}}
\label{subsec:observation_after_flare}

\begin{figure}[t]
    \centering
    \includegraphics[width=\hsize]{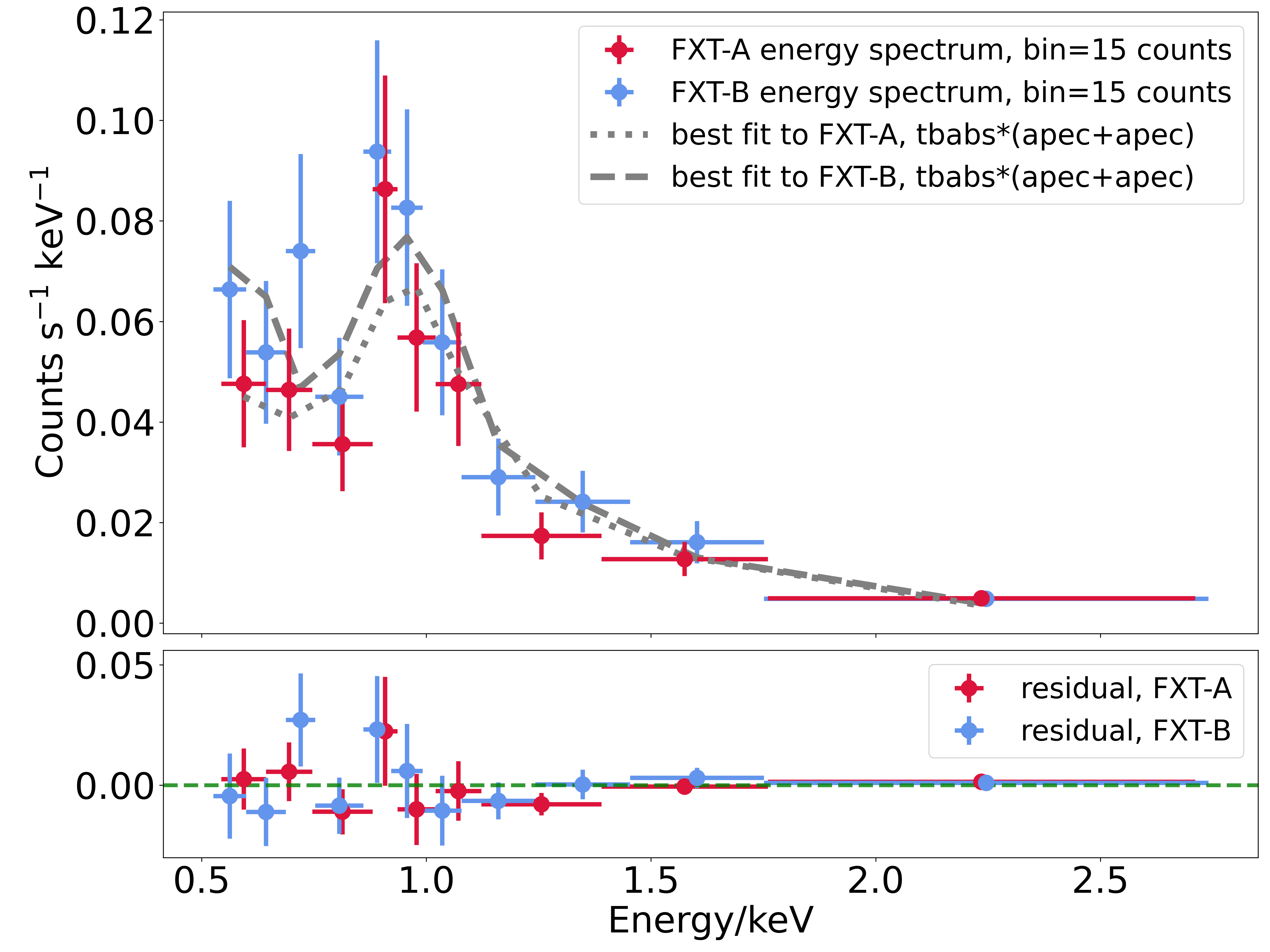}
    \caption{Best fit {and residuals} to the FXT energy spectrum of BP~Tau observed $\sim1.5$ day after the flare. Spectra from the two modules of FXT (named FXT-A and FXT-B) were fitted simultaneously, and are represented by the red and blue error bars (observation) and by the dotted and dashed curves (best-fit result), respectively. {The spectra are grouped with 15 photon counts per bin, and the error bars in the energy coordinate represent the energy coverage of each bin.}}
    \label{fig:fit_FXT_spec}
\end{figure}

A follow-up observation by FXT began $\sim1.5$ days after the observation by WXT, finding that BP~Tau dropped to its quiescent flux. The flux from FXT was calculated {in XSPEC \citep{Arnaud1996_xspec}} by fitting the energy spectrum using the Astrophysical Plasma Emission Code model \citep[APEC;][]{Smith2001_APEC} and the absorption model \texttt{tbabs} \citep{Wilms2000_tbabs}. Two plasma components subject to the same absorption were used in the fit, with the X-ray absorption fixed to $n_{\texttt{H}}=1.5\times10^{21}\texttt{cm}^{-2}$ as adopted in \citet{Robrade2006}. {Considering that both the abundance and plasma temperature influence the line-to-continuum ratio of the spectrum and the current data (especially WXT data) are not sufficient to break the degeneracy, simultaneously fitting both parameters prevents robust constraints. Hence, we froze the abundance to 0.1 times solar abundance, based on the measured range of 0.1–0.2 for iron in TW Hya \citep{Brickhouse2010}, which displays a similar high resolution spectrum to BP Tau \citep{Schmitt2005}.} The best-fit result is shown in Fig. \ref{fig:fit_FXT_spec}. The flux between 0.5 and 10.0 keV is 
{$(4.6^{+0.2}_{-0.5})\times10^{-13}$} erg cm$^{-2}$ s$^{-1}$, within 20\% of the 0.3-10.0 keV flux in \citet{Robrade2006} ($5.6\times10^{-13}$ erg cm$^{-2}$ s$^{-1}$, converted from the luminosity in their Table 3 and a distance of 140 pc adopted by \citealp{Robrade2006}). During the time between the WXT and FXT observation when BP~Tau was affected by the gap between TESS detectors after the flare, the ASAS-SN light curve shows a flux decay, albeit with lower time resolution than TESS.

\begin{table}[t]
    \caption{Timescales and energies of the X-ray and optical flares.}
    \label{table:flare_fitted_paras}
    \centering    
    \begin{threeparttable}
    \begin{tabular}{lllll}
    \hline
    \hline
           & WXT   & TESS   \\ \hline
    $\rm t_{peak}$ [MJD-60650.073] & 0.029\tnote{(a)} & 0.029   \\
    $\tau_{\rm rise}$ [ks]   & $1.7\pm1.0$ & $0.34\pm0.04$  \\
    $\tau_{\rm decay}$ [ks]  & $14\pm5$ & $0.34\pm0.04$\tnote{(b)}  \\
    Energy [$10^{34}$ erg] & $100\pm20$ & $2.8\pm0.4$ \tnote{(c)} &  \\ \hline
    \end{tabular}
    \begin{tablenotes}
        \item[(a)] As WXT did not detect the peak of the flare due to obstruction by the Earth, we assume that both the X-ray and optical flares reached the peak at the same time.
        \item[(b)] The decay time of the optical flare is set equal to the rise time, and both phases were fitted simultaneously due to the few data points in the decay phase.
        \item[(c)] The TESS energy corresponds to a bolometric energy of $1.9\pm0.3\times10^{35}$ erg, assuming an 11000 K blackbody spectrum for the optical flare.
        \end{tablenotes}
    \end{threeparttable}
\end{table}

\subsection{Flare timescale and energy}
\label{subsec:flare_energy}

Timescales and energies {of the X-ray and optical flares are} presented in Table \ref{table:flare_fitted_paras}. We estimated the X-ray flare timescale and energy by fitting the observed light curve with an exponential rise and decay model (Eq.~B1 in \citealp{Getman2021b}). Parameters in the fit were constrained by fixing the peak time to the peak of the optical flare. While a time lag between optical and X-ray observations is found in some flares \citep[e.g.,][]{Stelzer2006, Stelzer2022b}, such a lag lasts only a few minutes and does not significantly affect the fit result. Additionally, we restricted the flare e-folding rise ($\tau_{\mathrm{rise}}$) and decay times ($\tau_{\mathrm{decay}}$) to within $\pm0.5$ dex of the relation in \citet{Getman2021b}, where we adopted local regression fits to the data in their Tables 1 and 2 to obtain the relation and 0.5 dex corresponds to the 1$\sigma$ uncertainty. Hence, the fit includes three free parameters, a peak count rate, $\tau_{\mathrm{rise}}$, and deviations derived from the relation in \citet{Getman2021b}. These are $0.016\pm0.005$ counts s$^{-1}$, $1.7\pm1.0$ ks, and $0.26\pm0.16$ (in units of $\log(\mathrm{t/ks})$; see Fig. \ref{fig:alignment_WXT_TESS} for the best-fit result). The corresponding decay time $\tau_{\mathrm{decay}}$ is $14\pm5$ ks.

\begin{figure}[t]
    \centering
    \includegraphics[width=\hsize]{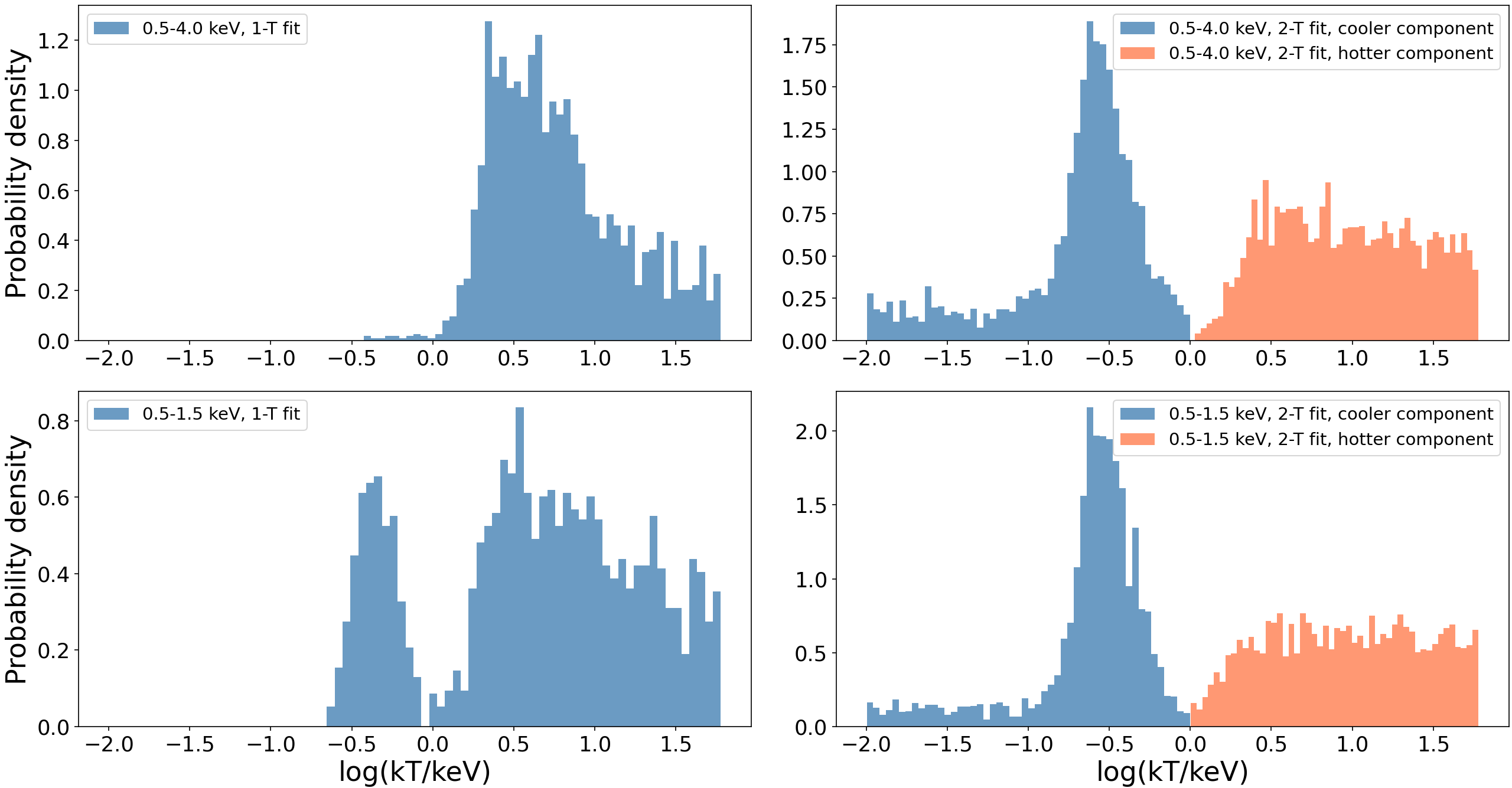}
    \caption{{Posterior distribution of $\log kT$ in the fit to the WXT energy spectrum with different settings: 0.5-4.0 keV energy range and one-component model (top left); 0.5-4.0 keV energy range and two-component model (top right); 0.5-1.5 keV energy range and one-component model (bottom left); 0.5-1.5 keV energy range and two-component model (bottom right).}}
    \label{fig:compare_BXA_fit_kT_posterior}
\end{figure}

\begin{figure}[t]
    \centering
    \includegraphics[width=\hsize]{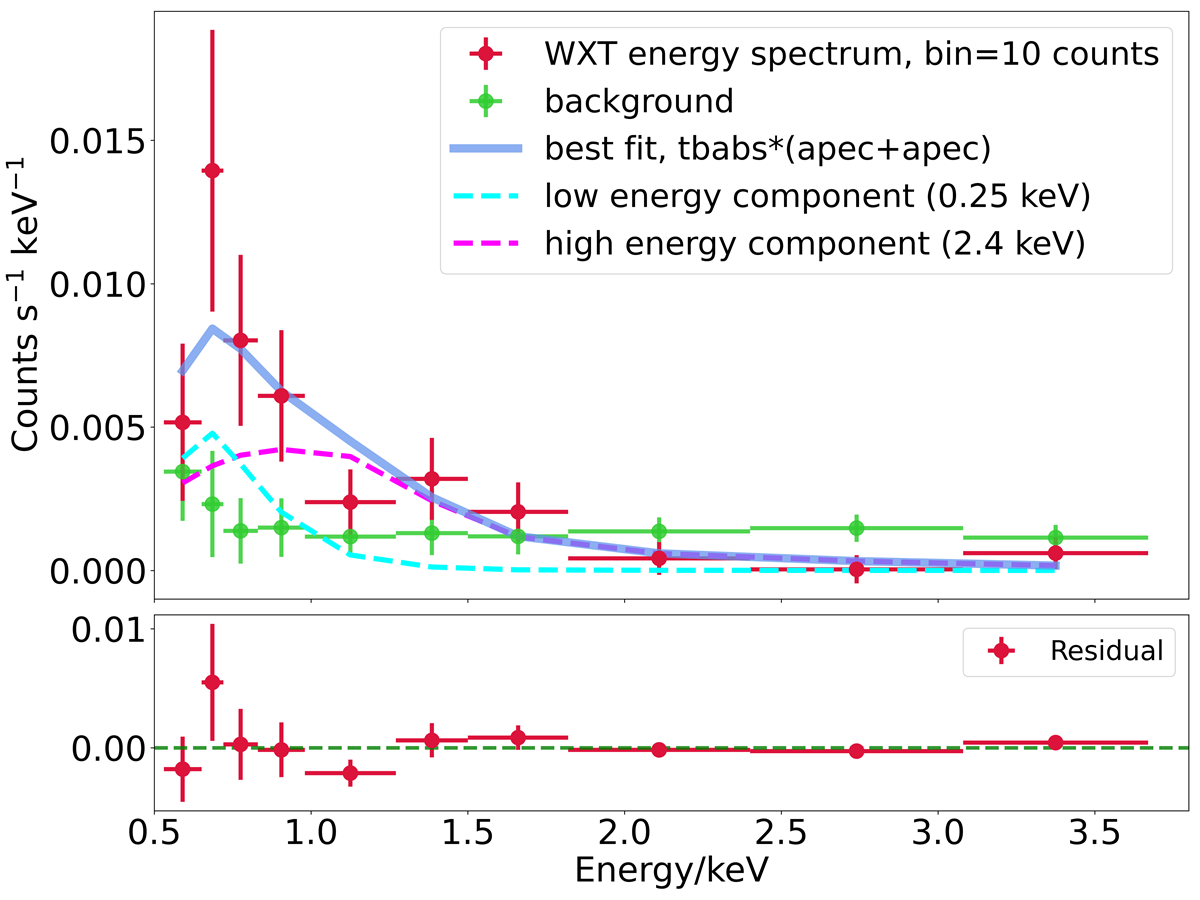}
    \caption{{Fit result} (blue curve) to the background-subtracted WXT energy spectrum of BP~Tau's flare (red error bars). {The contributions from the cool and hot components are shown with the dashed curves. {The background spectrum is represented by the green error bars.} The spectra are grouped with ten photon counts per bin, and the error bars in the energy coordinate represent the energy coverage of each bin.}}
    \label{fig:fit_WXT_spec}
\end{figure}

To convert the count rate to flux and calculate the flare energy, we fitted the average X-ray energy spectrum using {the Bayesian X-ray Analysis \citep{Buchner2014_BXA} software, which connects the nested sampling algorithm UltraNest \citep{Buchner2021_UltraNest} with the fitting environment XSPEC. The plasma model, absorption and metallicity were the same as those in the fit to the FXT spectra. We compared the difference in the log-Bayesian evidence between the single- and two-component plasma models ($\Delta \ln Z = \ln Z_{\mathrm{two}}-\ln Z_{\mathrm{single}}$) and evaluated whether the two-component model significantly improves the fit through simulations. In the single-component model fit, the $kT$ range was set to 0.01-60 keV, while in the two-component model fit, the $kT$ ranges were set to 0.01-1.0 and 1.0-60 keV, representing the cooler and hotter components. Figure \ref{fig:compare_BXA_fit_kT_posterior} presents the posterior distribution of the fitted $kT$. The single-component fit exhibits a peak at $kT\sim3.2$ keV. The two-component fit displays a peak at $\sim0.25$ keV and cannot constrain the hotter component, with a lower limit of 2.4 keV (90\% confidence level). The difference in the log-Bayesian evidence $\Delta \ln Z$ is 1.7. Then, we simulated 200 spectra via XSPEC based on the single-component model, the observation's Response Matrix File, Ancillary Response File, background spectrum, and Poisson errors. The $kT$ in the model was set to 0.25 keV, which corresponds to the peak in the two-component fit. Each generated spectrum was fitted with both single-component model and two-component model, leading to the distribution of $\Delta \ln Z$. Among these simulations, four $\Delta \ln Z$ values are greater than 1.7, resulting in a fraction of 2\%. We also performed the same analysis using the energy spectrum in the 0.5-1.5 keV range to mitigate background contamination, as the background flux surpasses the net flux at higher energies (see Fig. \ref{fig:fit_WXT_spec}). The $kT$ posterior distribution exhibits two peaks at $\sim0.4$ and $\sim3.2$ keV in the single-component fit, as shown in Fig. \ref{fig:compare_BXA_fit_kT_posterior}, and the $\Delta \ln Z$ is 1.9. In the simulation, the model $kT$ used to generate the simulated energy spectra was set to 0.4 keV, corresponding to the peak with a lower temperature in the posterior distribution of the single-component fit. Three simulated $\Delta \ln Z$ values were greater than 1.9, leading to a fraction of 1.5\%. Hence, the improvement in the fit by adding a high-temperature component is unlikely (2\% probability) to be caused by the single-component model combined with Poisson fluctuations.}

{Applying the two-component model to the fit, the contribution from the two components to the observation is shown in Fig. \ref{fig:fit_WXT_spec}, in which the cooler component is set to 0.25 keV (the median of its posterior distribution) and the hotter component is set to 2.4 keV (the lower limit). Despite the unconstrained upper limit of the hot component's temperature, the posterior distribution of the fitted flux converges, with a median and 68\% confidence interval of $(1.5^{+0.3}_{-0.4})\times10^{-11}$ erg cm$^{-2}$ s$^{-1}$.} {Converting the count to the flux with a factor of $2.3\times10^{-9}$ based on the average count rate and flux,} the X-ray flare peak flux is $(3.7\pm1.2)\times10^{-11}$ erg cm$^{-2}$ s$^{-1}$, {$\sim8$ times the brightest peak flux of previously detected BP~Tau's flares}\footnote{{{Chandra} ObsID 16205. The peak count rate is $\sim8$ times the quiescent level, while the fitted peak flux of the WXT light curve is $\sim62$ times the quiescent flux of BP~Tau in \citet{Robrade2006}.}}. The peak flux corresponds to a peak luminosity of $(7\pm2)\times10^{31}$ erg s$^{-1}$. Integrating the fitted count rate over time leads to a total of $230\pm50$ counts, corresponding to a total flare energy between 0.5 and 4.0 keV of $(1.0\pm0.2)\times10^{36}$ erg. The quiescent X-ray flux of BP~Tau in \citet{Robrade2006} is $\sim4\%$ times the flux observed here and contributes little to the WXT observation. This event qualifies as a superflare according to the classification in \citet{Getman2021a}.

Comparing with the large sample of T Tauri X-ray superflares in \citet{Getman2021a}, this flare falls near the 40th percentile of the cumulative distribution function of the flare peak luminosities and near the median of the cumulative distribution function of flare energies, as shown in Fig. 5 of \citet{Getman2021a}. Adopting 0.3 day as the duration of this flare, this duration is near the 30th percentile of the cumulative distribution function of the flare durations in Fig. 7 of \citet{Getman2021a}.

Calculating the timescale and energy of the optical flare requires the complete shape of the flare, which is a mix of two peaks and some minor structures in the light curve. We modeled the {slow modulation baseline} as a three-degree polynomial fit to the data points near the flare and fitted the flare with an exponential rise and decay model. Due to the lack of data points in the decay phase, the decay time was set to be the same as the rise time. As the typical shape of the decay phase in an optical flare is a fast decay followed by a slow decay, our approximation models the fast decay phase while ignoring the slow decay. The peak time was fixed to 0.0292 day in the fit. The fitted peak flux is $0.076\pm0.006$ (relative to the flux of BP~Tau), and the e-folding rise time is $0.33\pm0.04$ ks. The best-fit results to the flare and the baseline at the time of the flare are shown in Fig. \ref{fig:alignment_WXT_TESS}.

The optical flare energy was subsequently calculated following the method in \cite{Davenport2016}, multiplying the equivalent duration of the flare by the luminosity of BP~Tau in the TESS band. The equivalent duration was calculated using $\int[F(t)/F_0-1]dt$, and is found to be $0.050\pm0.007$ ks. The TESS luminosity of BP~Tau, converted from $T_\mathrm{mag}=10.64$, the TESS filter zero point in the Vega system, and effective width calculated by the SVO Filter Profile Service \citep{Rodrigo2012_SVO, Rodrigo2020_SVO, Rodrigo2024_SVO}, is $5.6\times10^{32}$ erg s$^{-1}$. Multiplying the equivalent duration of the flare by the TESS luminosity of BP~Tau, the optical flare energy in the TESS band is $(2.8\pm0.4)\times10^{34}$ erg. This energy is consistent with previously observed TESS flares of BP~Tau, DN Tau, and FN Tau ($10^{34}$ to $10^{35}$ erg, \citealt{Lin2023}), with a duration shorter than those flares, which however ignores the long decay phase.

To calculate the bolometric energy of the optical flare, we assumed a 11\,000\,K blackbody emission to represent the flare, as in \citet{Jackman2023} where optical and UV observations were combined to estimate the flare temperature of M dwarfs. Integrating the 11\,000\,K blackbody spectrum over wavelength, the bolometric energy of the optical flare is 6.8 times the energy in the TESS band, corresponding to $(1.9\pm0.3)\times10^{35}$ erg. {This estimate is regarded as a lower limit. Assuming that the decay time equals the rise time leads to an underestimation of the energy, as the energy of the rise phase generally occupies $\sim20$\% of the total flare energy \citep{Davenport2014}. Extinction also reduces the estimated energy. The $A_V=0.45$ mag of BP~Tau \citep{Herczeg2014} corresponds to $A_{TESS}=0.3$ by adopting $A_{TESS}=2.06E(B-V)$ \citep{Stassun2019} and $A_V=3.1E(B-V)$. Accounting for these two factors, the estimated flare energy would increase by a factor of $\sim3$. Meanwhile, the current choice of the baseline under the flare merely isolates the peak. If other segments of the light curve also contribute to the flare, the estimated energy will also be higher.}

The energy of the {flux bump between $\sim-0.1<\mathrm{MJD}-60650.073<\sim0.3$ in the TESS band}, calculated by integrating over the individual data points, is $1.33\times 10^{36}$ erg, significantly higher than the energy of the flare.

\section{Discussion}
\label{sec:discussion}

As discussed in Sect. \ref{subsec:lc_interpret}, the TESS light curve consists of a flare and an accretion burst. Some observations and simulations suggest that flares may trigger accretion bursts in classical T-Tauri stars \citep[e.g.,][]{Orlando2011, Reale2018, Colombo2019, Espaillat2019}. Here, we consider whether the TESS light curve is consistent with this scenario. A visual inspection of the light curve reveals that the start time of {peak ''b'' in Fig. \ref{fig:lc_ASASSN_TESS_EP}} seems earlier than the flare. The time lag between the peak ''b'' and the flare is too short for a flare-triggered accretion burst (see the estimate below), also disfavoring the interpretation that peak ''b'' was triggered by the flare. Peak ''c,'' if isolated from the second, would be a more plausible candidate for a flare-triggered event. We estimated the probability of a statistical or random coincidence and compared the observed and estimated time lag between a flare and a flare-triggered accretion burst.

Assuming that the occurrence of flares and accretion bursts are independent, with a frequency of $f_1$ and $f_2$ and durations of $t_1$ and $t_2$, respectively, the probability that the two phenomena occur together is
\begin{equation}
    P=f_1t_2+f_2t_1-(f_1t_2)(f_2t_1).
\end{equation}
We adopted $f_1$ from Fig. 8 in \citet{Getman2021a}, i.e., ten flares per star per year for a flare energy $>10^{35}$ erg to account for the uncertainty in estimating the X-ray energy of BP~Tau's flare, and adopted a typical flare timescale of 50\,ks \citep{Getman2021a} as $t_1$. For the optical flare, we used 0.2 times per day and 0.3 day as $f_2$ and $t_2$ and refered to \citet{Stauffer2014} for the estimate of frequency and timescale. With these numbers, the probability of chance coincidence is 12\%; hence, we cannot rule out that the observed TESS variability represents a chance alignment of a flare and an independent accretion burst.

To estimate the time lag between a flare and a flare-triggered accretion burst, we examined a scenario in which a flare disturbs the inner disk, triggering disk material falling onto the stellar surface \citep[e.g.,][]{Orlando2011}.

In the commonly adopted protoplanetary disk model, the disk is truncated by the stellar magnetic field at a radius, $R_\mathrm{in}$, where the magnetic pressure roughly equals the gas ram pressure \citep[e.g.,][]{Koenigl1991, Hartmann2016}. The magnetic pressure can be expressed as $P_\mathrm{mag}=B^2/8\pi=\mu_*^2/4\pi R_{\mathrm in}^6$, where $\mu$ is the stellar magnetic moment. The ram pressure is $P_\mathrm{ram}\approx \rho v_\mathrm{ff,in}^2$, where the density $\rho=\dot{M}/4\pi R_\mathrm{in}^2 v_\mathrm{ff,in}$, and the free-fall velocity at the disk inner radius $v_\mathrm{ff,in}=(2GM_*/R_\mathrm{in})^{1/2}$ ($M_*$ and $\dot{M}$ denote stellar mass and accretion rate, respectively). Adopting $P_\mathrm{mag} \approx P_\mathrm{ram}$, the inner radius is
\begin{align}
    R_\mathrm{in}&=\xi\left(\frac{\mu_*^4}{2GM_*\dot{M}^2}\right)^{1/7}=\xi\left(\frac{B_*^4R_*^{12}}{4GM_*\dot{M}^2}\right)^{1/7} \\
    &=2.26\times10^{11}\xi B_3^{4/7}\dot{M}_8^{-2/7}m_*^{-1/7}r_*^{12/7}\ {\rm cm},
\end{align}
where $B_3=B_*/1\ {\rm kG}$, $m_*=M_*/M_\odot$, $r_*=R_*/R_\odot$, $\dot{M}_8=\dot{M}/10^{-8}M_\odot\ {\rm yr^{-1}}$ and $\xi$ is the correction factor ($\lesssim1$, depending on the interaction between the disk and magnetic field; \citealp{Hartmann2016}).

The time lag between the flare and accretion burst is approximately the free-fall timescale from the inner disk to stellar surface,
\begin{equation}
    \Delta t\approx\frac{R_\mathrm{in}-R_*}{v_\mathrm{ff,*}},
\end{equation}
where the free-fall velocity at the surface of star is $v_\mathrm{ff,*}\approx(2GM_*/R_*)^{1/2}$, ignoring the gravitational potential at $r=R_\mathrm{in}$ as $R_\mathrm{in}$ is several times larger than stellar radius.

For BP~Tau, we adopted $M_*=0.8M_\odot$, $R_*=2.1R_\odot$, $\dot{M}=2.9\times10^{-8}\ M_{\odot}\ \texttt{yr}^{-1}$ (all from \citealp{Ingleby2013}), and $B=2.5 kG$ \citep{Flores2019}\footnote{\citet{Donati2008} obtained a magnetic field strength of 1.2 kG. This difference likely reflects the complex morphology of the magnetic field.}. The correction factor was set to $\xi=0.7$, as in \citet{Hartmann2016}. Hence, the derived disk inner radius is $R_\mathrm{in}=7.2\times10^{11}$ cm ($\sim4.8R_*$, in agreement with previous estimates, e.g., $6R_*$ in \citealt{Long2005}; $>4R_*$ in \citealt{Donati2008}), and the time lag is $\Delta t\approx15.0$ ks.

We also derived another estimate, independent of the stellar magnetic field and accretion rate. Given that flares cause fluctuation in magnetic pressure ($\Delta P_\mathrm{mag}$), a flare disrupts the balance between magnetic pressure and gas ram pressure. Assuming that the X-ray flare energy is mainly contributed by the magnetic pressure fluctuation that results in disk material with mass, $m$, falling onto stellar surface, the released energy is
\begin{equation}
    E_\mathrm{X}\approx \int2\Delta P_\mathrm{mag}dV=\frac{4GM_*m}{R_\mathrm{in}}.
\end{equation}
The accretion burst results from the gravitational energy of the material falling onto the stellar surface due to the flare, and we used the optical accretion burst energy to approximate the energy release:
\begin{equation}
    E_\mathrm{opt}\approx\frac{GM_*m}{R_*}\left(1-\frac{R_*}{R_\mathrm{in}}\right).
\end{equation}
Therefore, we derived the time lag from the ratio between the flare energy and the accretion burst energy,
\begin{equation}
    \Delta t\approx\frac{R_\mathrm{in}-R_*}{v_\mathrm{ff,*}}=\frac{E_\mathrm{opt}}{E_\mathrm{X}}\frac{4R_*}{v_\mathrm{ff,*}}=4.5\frac{E_\mathrm{opt}}{E_\mathrm{X}}m_*^{-1/2}r_*^{3/2}\ {\rm ks},
\end{equation}
which is $\Delta t\approx15.2(E_\mathrm{opt}/E_\mathrm{X})$ ks. Adopting $E_\mathrm{opt}/E_\mathrm{X}=1$, as estimated in Sect. \ref{subsec:flare_energy}, the time lag is $\Delta t \approx15.3$ ks. This value is similar to the previous estimate. Such a timescale significantly exceeds the interval between {peak ''a'' and ''b''} in the TESS light curve and is closer to the $\sim10$ ks interval between {peak ''a'' and ''c''} .

This time lag estimate requires that the flare magnetic loop reaches the inner disk \citep[e.g.,][]{Favata2005, Aarnio2010}. We estimated the loop height, $L$, following the approach in \citet{Mao2025}, using the conductive loss cooling timescale and neglecting the difference between the maximum loop temperature and the averaged loop temperature, with a flare loop height (expressed in centimeter-gram-second):
\begin{equation}
    L=1.9\times10^{3}\tau_\mathrm{decay} T^{1/2}.
\end{equation}
Adopting $\tau_\mathrm{decay}=14$ ks from the fit in Sect. \ref{subsec:flare_energy} and $T=$ 5 keV, the loop height reaches an order-of-magnitude of $10^{11}$ cm, a value similar to the disk inner radius, indicating the possibility that the flare loop reaches the inner edge of the disk. More observational data on such phenomena will further constrain the nature of the TESS light curve.

\section{Conclusions and prospects}
\label{sec:conclusion}

A superflare in BP~Tau was simultaneously observed by WXT and TESS, with flare energies of $(1.0\pm0.2)\times10^{36}$ erg (0.5-4.0 keV) and $(2.8\pm0.4)\times10^{34}$ erg (TESS band). We estimated the TESS flare energy by setting the decay timescale equal to the rise timescale due to the lack of data points in the decay phase. The energy of the optical flare corresponds to a bolometric energy of $(1.9\pm0.3)\times10^{35}$ erg assuming a 11\,000\,K blackbody spectrum. The X-ray flare has e-folding times of $1.7\pm1.0$ ks and $14\pm5$ ks for the rise and decay phases, and the optical flare has an e-folding time of $0.34\pm0.04$ ks for the rise phase. The energy and timescale of the X-ray flare make it a superflare when compared with previous large-sample observations. The energy of the optical flare is consistent with previously observed optical flares on classical T-Tauri stars. The timescale appears to be shorter due to the underestimation of the decay timescale.  

The morphology of the whole burst {between $\sim-0.1<\mathrm{MJD}-60650.073<\sim0.3$} in the TESS light curve is explained by a flare followed by an accretion burst, which suggests the possibility that a flare can trigger an accretion burst. Given the frequency of both phenomena, we cannot rule out the possibility of a coincidental event. Stronger evidence for the flare-triggered accretion burst requires additional observations and statistical data.

This work illustrates the potential of using the large field of view of WXT and TESS for flare analysis. During the $\sim20$ months of EP in-orbit observations (including $\sim7$ months commissioning and $\sim13$ months nominal operation), WXT detected $\sim1400$ flares, corresponding to $\sim800$ flares per year and demonstrated the ability to study flares, such as a those with potential chromospheric evaporation \citep{Wang2024} and the most energetic flare \citep{Gunther2024, Mao2025} to date. Assuming a 3 ks observation (the typical unblocked observing window of one orbit) and that the detection limit is inversely proportional to the observation duration, the limiting flux of WXT is $\sim8.7\times10^{-12}$ erg s$^{-1}$ cm$^{-2}$. If a star at 60 pc flares and is detected by WXT, the averaged flare luminosity is $\sim3.7\times10^{30}$ erg s$^{-1}$, placing it among the most luminous flares (e.g., peak luminosity exceeding $3.2\times10^{30}$ erg s$^{-1}$, \citealt{Getman2021a}). Hence, most flares caught by WXT are likely to be superflares, and some may be mega-flares. Given the rarity of super- and mega-flares, data accumulated through WXT sky surveys will yield a significant and possibly unprecedented sample of such events.

WXT's large field of view and high scanning cadence ($\sim100$ minutes per orbit and three orbits covering the night sky, \citealt{Yuan2025_EP})  make it possible to point toward the same sky area together with TESS. Considering the observational strategy of TESS, stars with high ecliptic latitudes are more frequently covered. Pre-main-sequence stars, for example, in the Chamaeleon molecular clouds and some regions in the Scorpius-Centaurus OB association are located at higher ecliptic latitude and are nearby ($<200$ pc, \citealt{Galli2021, Luhman2022}), making them ideal targets for flares caught by both WXT and TESS.

In addition to simultaneous observations, X-ray events captured by WXT may also trigger follow-up of ground-based telescopes, resulting in quasisimultaneous light curves. FXT reaches deeper sensitivity and can be automatically triggered for follow-up observations $\sim3$-5 minutes after WXT detects an X-ray event \citep[e.g.,][]{Jiang2025, Yin2025}. If a flare triggers an immediate follow-up, the data will better constrain flare properties such as X-ray absorption and plasma temperature near the flare's peak time and clearly show the decay phase following the peak. This may reveal whether a flare would trigger an accretion burst \citep[e.g., an accretion burst may cause increasing X-ray absorption,][]{Guarcello2017}. With more data being acquired, (quasi-)simultaneous X-ray and optical observation will detect more unique flares as well as provide a statistical flare sample.

\begin{acknowledgements}
We thank the anonymous referee for the constructive comments and suggestions that significantly improve the analysis and clarity of the paper. XYZ is supported by the Shanghai Sailing Program (Grant No. 24YF2750800). PCS is supported by the grant DLR 50 OR 2412. This work is based on data obtained with the Einstein Probe, a space mission supported by the Strategic Priority Program on Space Science of the Chinese Academy of Sciences, in collaboration with ESA, MPE, and CNES. This paper includes data collected by the TESS mission, which are publicly available from the Mikulski Archive for Space Telescopes (MAST). Funding for the TESS mission is provided by NASA's Science Mission directorate. We acknowledge the use of TESS High Level Science Products (HLSP) produced by the Quick-Look Pipeline (QLP) at the TESS Science Office at MIT, which are publicly available from the Mikulski Archive for Space Telescopes (MAST). This paper makes use of AstroImageJ, an astronomical image analysis software package based on ImageJ. This work presents results from the European Space Agency (ESA) space mission Gaia. Gaia data are being processed by the Gaia Data Processing and Analysis Consortium (DPAC). Funding for the DPAC is provided by national institutions, in particular the institutions participating in the Gaia MultiLateral Agreement (MLA). The Gaia mission website is https://www.cosmos.esa.int/gaia. The Gaia archive website is https://archives.esac.esa.int/gaia. This research has made use of the Spanish Virtual Observatory (https://svo.cab.inta-csic.es) project funded by MCIN/AEI/10.13039/501100011033/ through grant PID2020-112949GB-I00. 
\end{acknowledgements}

\bibliography{export-bibtex}

\end{document}